\documentclass[onecolumn, numberedappendix]{aastex6}
\usepackage{graphicx}
\usepackage{epsfig}
\usepackage{natbib}
\usepackage{amssymb}
\usepackage{amsbsy}
\usepackage{natbib}
\usepackage{url}
\usepackage[mathcal]{euscript}
\usepackage{hyperref}
\bibpunct{(}{)}{,}{a}{}{,}

\newcommand{\etal}{et~al.}

\begin{document}
 
\def\simlt{\vcenter{\hbox{$<$}\offinterlineskip\hbox{$\sim$}}}
\def\simgt{\vcenter{\hbox{$>$}\offinterlineskip\hbox{$\sim$}}}
\def\etal{et al.\ }
\def\kms{km s$^{-1}$}

\title{The Science Advantage of a Redder Filter for WFIRST}
\author{John Stauffer\altaffilmark{1}, 
George Helou\altaffilmark{2},
Robert A. Benjamin\altaffilmark{3},
Massimo Marengo\altaffilmark{4}, 
J. Davy Kirkpatrick\altaffilmark{2},
Peter Capak\altaffilmark{2},
Mansi Kasliwal\altaffilmark{5},
James M. Bauer\altaffilmark{6},
Dante Minniti\altaffilmark{10,11,12},
John Bally\altaffilmark{13},
Nicolas Lodieu\altaffilmark{7},
Brendan Bowler\altaffilmark{8},
ZengHua Zhang\altaffilmark{9},
Sean J. Carey\altaffilmark{1},
Stefanie Milam\altaffilmark{14},
Bryan Holler\altaffilmark{15}
}
\altaffiltext{1}{IPAC, MS 314-6, Caltech, 1200 E. California Blvd., Pasadena, CA 91125, USA: stauffer@ipac.caltech.edu}
\altaffiltext{2}{IPAC, MS 100-22, Caltech, 1200 E. California Blvd., Pasadena, CA 91125 USA}
\altaffiltext{3}{Department of Physics, University of Wisconsin-Whitewater, Whitewater, WI, USA}
\altaffiltext{4}{Department of Physics and Astronomy, Iowa State University, Ames, IA 50011}
\altaffiltext{5}{Astronomy Department, California Institute of Technology, Pasadena, CA 91125}
\altaffiltext{6}{University of Maryland, Department of Astronomy, College Park, MD 20742}
\altaffiltext{7}{Instituto de Astrof\'isica de Canarias, C. via Lactea s/n, 38200 La Laguna, Tenerife, Spain}
\altaffiltext{8}{McDonald Observatory and the Department of Astronomy, University of Texas at Austin, Austin, TX 78712}
\altaffiltext{9}{GEPI, Observatoire de Paris, PSL Research University, CNRS, 5 Place Jules Janssen, 92195 Meudon, France}
\altaffiltext{10}{Departamento de F\'isica, Facultad de Ciencias Exactas, Universidad Andr\'es Bello, Av. Fernandez Concha 700, Las Condes, Santiago, Chile.},
\altaffiltext{11}{Instituto Milenio de Astrof\'isica, Santiago, Chile.},
\altaffiltext{12}{Vatican Observatory, V00120 Vatican City State, Italy.},
\altaffiltext{13}{Astrophysical and Planetary Sciences Department, University of Colorado, UCB 389 Boulder, CO 80309, USA}
\altaffiltext{14}{NASA Goddard Space Flight Center, 8800 Greenbelt Road, Greenbelt, MD 20771, USA}
\altaffiltext{15}{Space Telescope Science Institute, 3700 San Martin Drive, Baltimore, MD 21218, USA}
\begin{abstract}

WFIRST will be capable of providing Hubble-quality imaging performance over 
several thousand square degrees of the sky.   The wide-area, high spatial resolution
survey data from WFIRST will be unsurpassed for many decades into the
future.  With the current baseline design, the WFIRST filter complement will
extend from the bluest wavelength allowed by the optical design to a reddest
filter (F184W) that  has a red cutoff at 2.0 microns.  In this white paper, we outline
some of the science advantages for adding a K$_s$\ filter ($\lambda_c$\ $\sim$\ 2.15 $\mu$m)
in order to extend the wavelength coverage for WFIRST as far to the red as the 
possible given the thermal performance of the observatory and the sensitivity
of the detectors.

\end{abstract}

\section{Introduction}

WFIRST was the top priority of the National Academy of Sciences' 2010
Decadal Survey for Astronomy and Astrophysics.  Its design has been optimized
to conduct groundbreaking studies of dark energy and exoplanet demographics.
Those dark energy and exoplanet research programs are enabled by WFIRST's wide-field
camera which will provide imaging with HST sensitivity and spatial resolution
over an instantaneous field of view that is nearly two orders of magnitude larger.
While the majority of the observing
time during WFIRST's prime mission will be devoted to 
the core dark energy and exoplanet microlensing surveys, about 25\% of the
observing time is being set aside for community-specified and peer-reviewed
Guest Observer (GO)
programs that  could address
any science topic.  The ability of WFIRST to address the most
diverse and exciting science at the time of the mission would be greatly enhanced by providing a long
wavelength filter that is centered as far to the red as allowed by the telescope
and detector design.   Compelling science enabled by such a red filter
include, but are not limited to:
\begin{itemize}

\item  Exploring galactic structure through kinematics of the inner disk
       and bar/bulge;
\item  Determining the fragmentation limit for star formation and
       detecting free-floating planets in nearby star-forming
       regions;
\item  Characterizing the substellar population of the
       thick disk and halo;
\item  Anchoring a $K$-band cosmic distance ladder to be
       constructed with JWST;
\item  Better characterizing the end points of stellar evolution for
       the most massive stars;
\item  Enabling a range of improved extragalactic studies for z $>$ 3
       galaxies, including the ability to identify 11 $<$ z $<$ 15 galaxies, should they exist;
\item  Identifying water ice on a sample of the most distant objects in
       the solar system.
\end{itemize}

We discuss each of these points in more detail in the
remainder of the document.

\section{Discovery Space Advantage of A $K$ Filter }

The deepest, wide-area $K$-band survey of the sky is the VISTA VHS survey (Banerji \etal\ 2015),
now nearing completion.  The VHS covers the entire southern sky to a 
depth of about $K$(Vega) = 18 mag, 5$\sigma$.  As illustrated in the Appendix,
a $K_s$\ filter on WFIRST could obtain survey data 
to a depth that is five magnitudes deeper than the VHS.    Using the same
exposure times and dithering strategy as for the planned WFIRST high-latitude
survey, the WFIRST survey speed will be about 20 square degrees per day per
filter.  No other existing or planned wide-area survey could come anywhere
close to imaging such large areas to these faint limits at this wavelength.

\begin{figure}[ht]
\includegraphics[scale=0.7,angle=0]{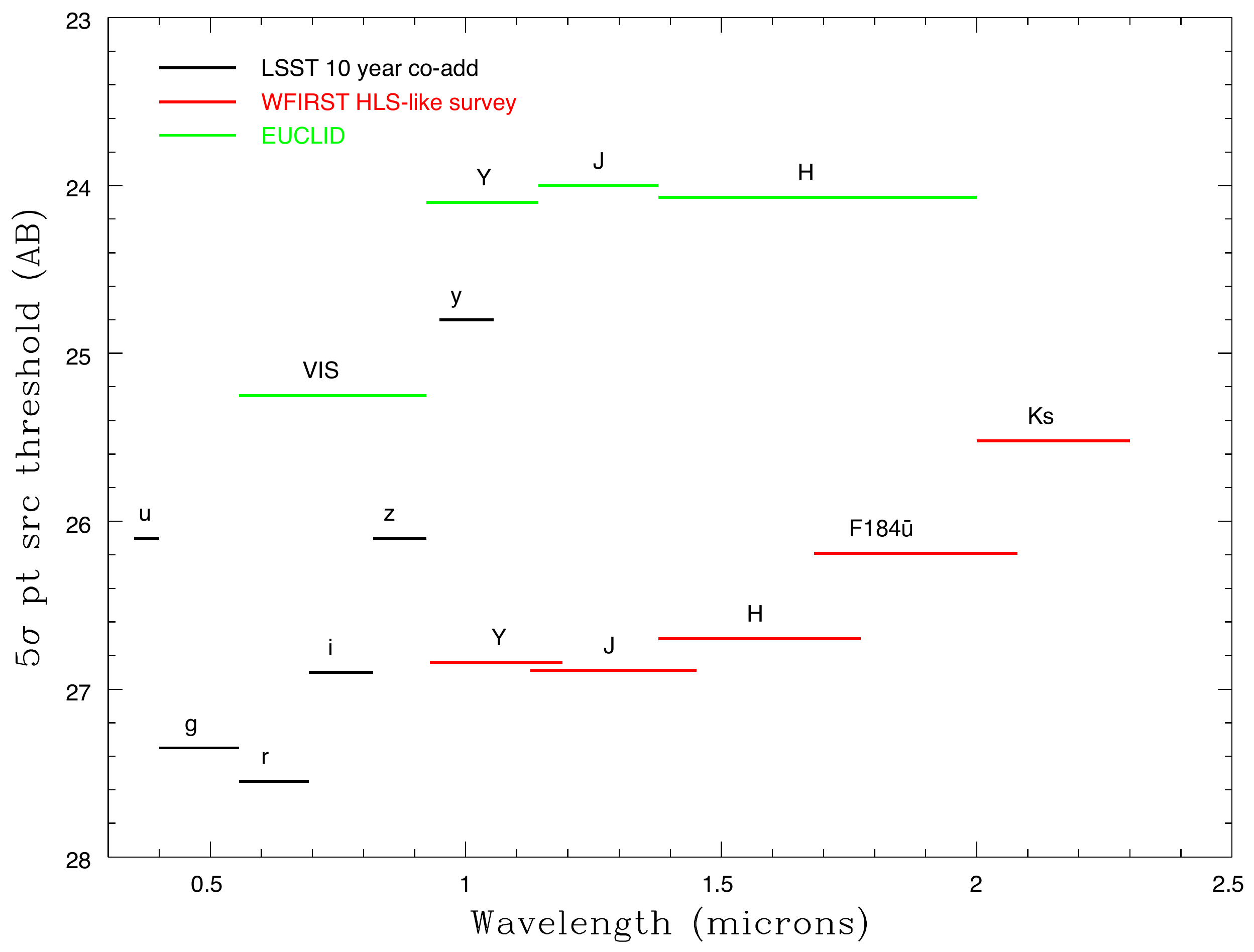}
\caption{Comparison of the sensitivities of WFIRST, EUCLID and LSST.
The LSST values are for the full 10 year co-add data at the end of the survey.
Depth in AB magnitudes.  WFIRST limits are for its High Latitude Survey (HLS).  
The current baseline for WFIRST, although not shown here,
includes a ``V" filter ($\lambda\lambda$0.45-0.7 $\mu$m) and a 
z-filter ($\lambda\lambda$0.76-0.98 $\mu$m).   In this white paper,
we advocate adding a filter longward of 2.0 microns,
similar in properties to the 2MASS K$_s$\ filter.
The sensitivity for our proposed K$_s$ filter is shown in the figure based on an
assumed telescope temperature of 260 K (the F184 filter sensitivity is also
for that assumed temperature).
\label{fig:WF_EUC_LSS}}
\end{figure}

There is no viable replacement for the kind of very
deep, wide-field $K$ band imaging that WFIRST could provide.  WFC3 on HST does
not go longward of $H$ band; neither does EUCLID.   Those facilities did
not avoid $K$ band because it was of little interest -- it was simply not
feasible to include because thermal emission from the optics would have badly
degraded performance at those wavelengths.
The baseline plan for WFIRST is that it will be operated at $<$\ 270K, cold enough
to allow efficient use of a $K$-band filter.
We should make the most of this opportunity of a cool, wide-field telescope
in space by providing WFIRST with the reddest possible filter.
This red-filter discovery potential is still great 
even if WFIRST is not natural-background-limited at K$_s$\ because  
the potential gain in sensitivity at K-band relative to the ground is so large.
All of
the benefits of a redder filter described here rely on a $\it{relatively}$\rm\
cold operating temperature and the stable PSF, small pixel size and accurate
photometry that will be available at L2 regardless of the exact operating temperature.
Moreover, the addition of a K$_s$\ filter would make WFIRST more consistent
with the Decadal recommendation for a {$\it{near-infrared}$} telescope.

\begin{figure}[ht]
\epsscale{0.9}
\plotone{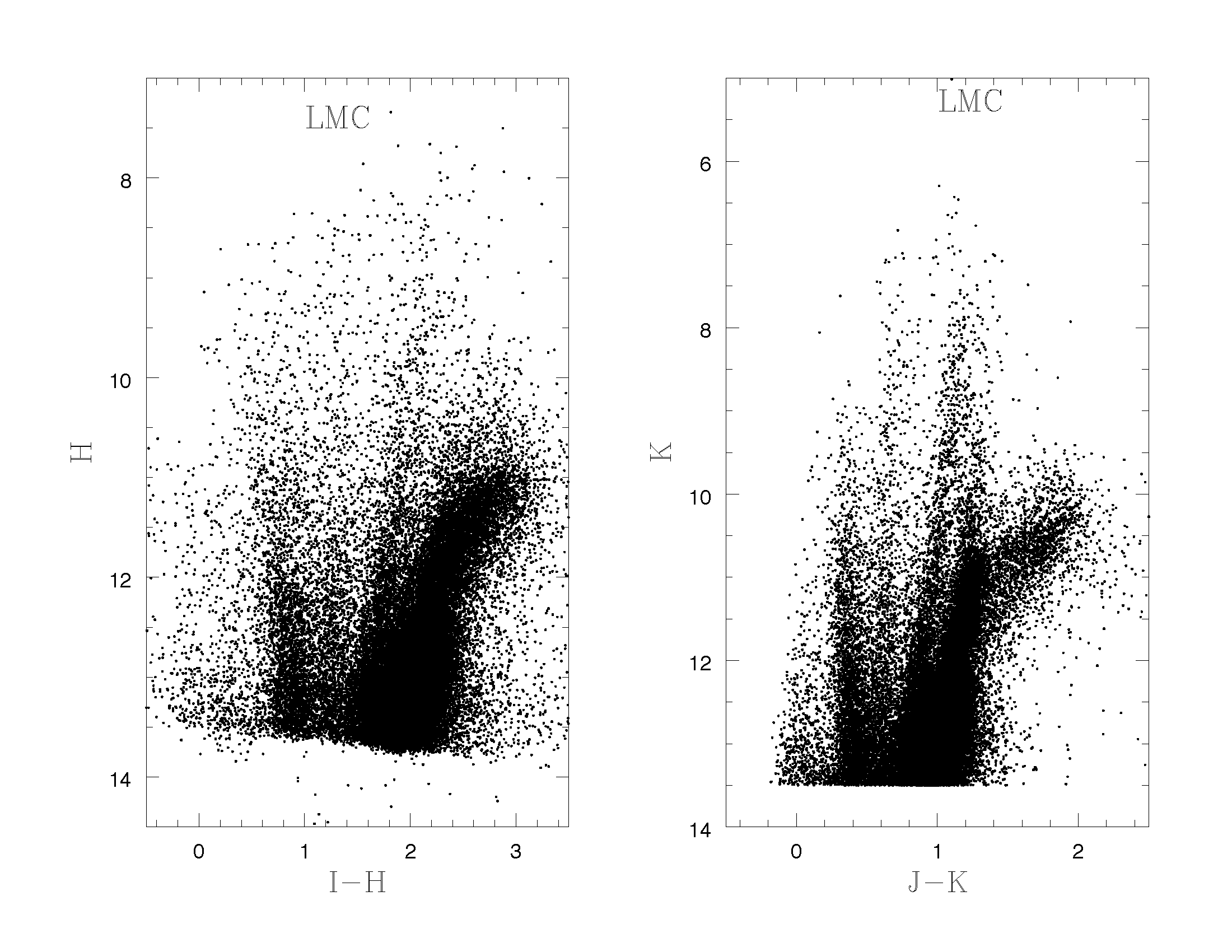}
\caption{Two versions of the color-magnitude diagram for giant stars
in the LMC, one utilizing $I-H$ for the $x$-axis and one using $J-K$.  
The same stars are plotted in the two panels. The $J-K$ diagram 
clearly distinguishes different sequences better than the former, particularly
for carbon stars (the finger pointing up and to the right for $J-K$ $>$ 1.4,
and oxygen rich supergiants (the narrow finger extending above the giant
branch, beginning at $J-K$ = 1.2 and $K$ = 10.5). The stars plotted are those
in the Spitzer SAGE (Meixner \etal\ 2006) LMC survey; we have cross-matched those
stars to the 2MASS and SDSS databases.  See Boyer \etal\ (2011) for a discussion
of the post-main sequence demographics that can be distinguished if a K
filter is available.
\label{fig:LMC_colormag}}
\end{figure}

There are simple astrophysical arguments why a redder filter would
open up important new regions of discovery space.
For giants, IR colors that include a $K$ band are optimal
for separating populations (see \citealt{nikolaev2000}
and Figure \ref{fig:LMC_colormag}).   The H$^-$\ opacity minimum (and
Rayleigh-Jeans spectra) results in stars with a very broad range of
spectral types and luminosities having very similar H-K colors; the range
from types B to K is less than 0.2 magnitudes.  This means that this
color can be used to estimate extinction when it is large, with relatively
small errors (see \S 3).  Similarly, the strong molecular bands in brown
dwarfs result in their J-K color providing a means to accurately estimate
their metallicity, making it possible for WFIRST to identify and characterize
the substellar population of the thick disk and halo better than any other
facility (see \S 5).  Finally, accurate distances to RR Lyrae stars (to
map the structure of tidal streams in the MW or as the first rung of a
Pop II distance ladder) is best done with WFIRST with a K filter - the 
Period-Luminsosity (PL)
relation is shallower (or non-existent) at shorter wavelengths and both 
extinction and metallicity effects are lower at K than at shorter bands.
We address specific
science enabled by a redder filter in more detail in the next several
sections.

\section{Structure and Kinematics of the MW Bulge, Nuclear Cluster and Inner Disk}

The inner 3 kpc region of our galaxy is much more complex
than once thought. The existence of a strong bar was first
recognized more than 25 years ago (\citealt{blitz1991}); Spitzer
IRAC data (\citealt{benjamin2005}) and ground-based NIR data 
(\citealt{wegg2015}) have recently provided much improved
maps of the extent and orientation of the bar. The
shape of the bulge is tri-axial rather than spheroidal, resulting
in an X-shaped or boxy appearance (\citealt{ness2016}). Most of the mass in
the bulge resides in a relatively metal rich (Fe/H $\sim$ $-$0.5 to 0),
rotationally supported population which traces the bar. However,
a minority ``hot'' (high RMS velocity, low rotation), more spheroidal, more
metal poor population (Fe/H $\sim$ $-$1) is also present 
(\citealt{dekany2013, gran2016}). Finally,
there is possible evidence for a young population interior to the 3 kpc
arm which may have more disk-like structure, possibly containing
up to 10\% of the bulge mass (\citealt{portail2016}).

\begin{figure}[ht]
\includegraphics[scale=0.65,angle=-90]{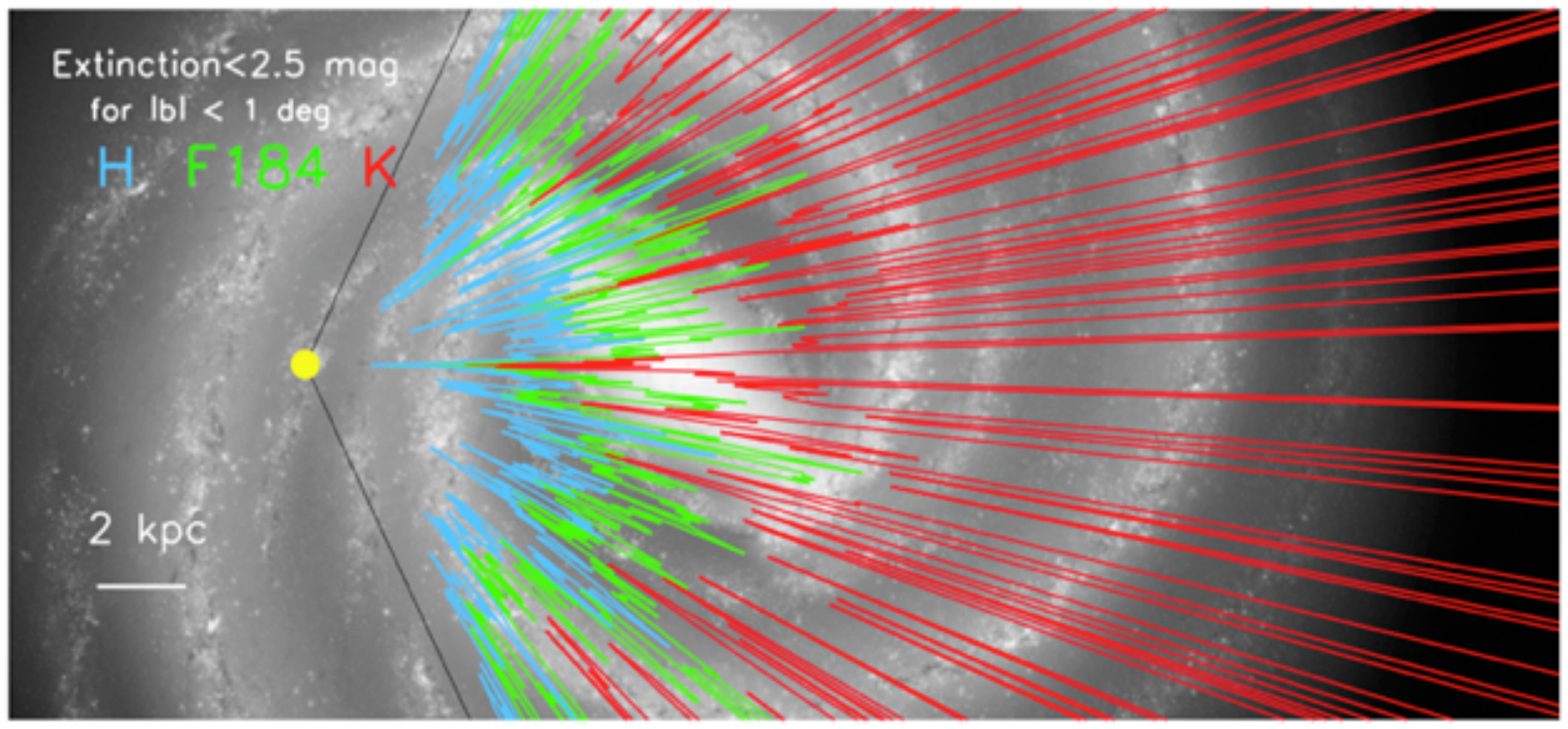}
\caption{
The extinction ``horizon" in the H, F184 and K band, i.e. the distance where 
$A_{band}>2.5$, as a function of Galactic longitude superimposed on an 
``artist's conception" of the Milky Way Galaxy by Robert Hurt. This uses 
the three-dimensional dust extinction model of Marshall et al (2006).  
At each longitude slice, we plot the distance corresponding to the maximum 
extinction at that distance for any latitude value. Note the volume of 
the Galaxy with modest (less than a factor of ten) extinction expands 
dramatically by going to the K band.
\label{fig:disk_horizon}}
\end{figure}

\begin{figure}[ht]
\epsscale{0.9}
\plotone{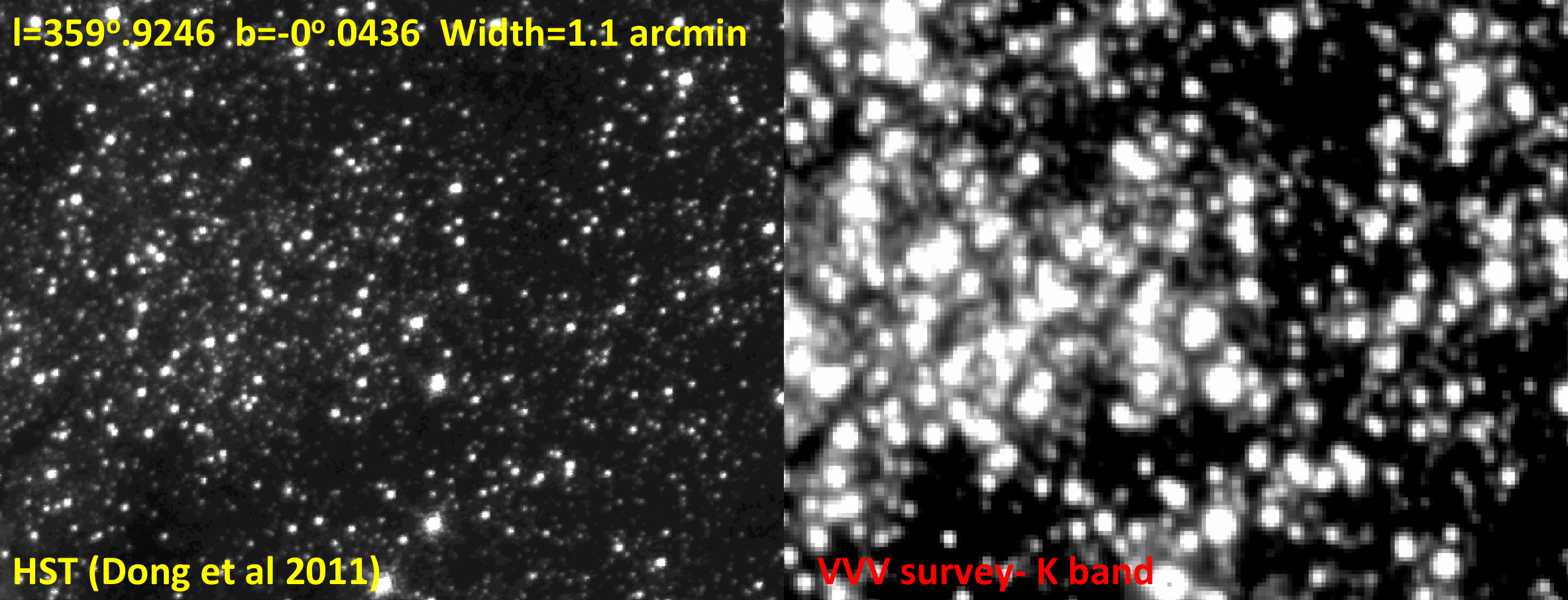}
\caption{
A comparison of a small pointing (1.1 arcmin in width) near the Galactic center. 
The left panel shows a HST F190 (1.9 $\mu m$) image from Dong et al (2011); 
the right panel shows a ground-based $K_{s}$ image from the VVV survey 
(Minniti et al 2010).  Note the dark areas in both images due to high 
extinction clouds toward the Galactic center. With a K-filter, WFIRST would 
have the best of both programs: the ability to see through a higher dust 
column in the K-band {\it combined} with spectacular angular resolution.
\label{fig:hst_VVV}}
\end{figure}

While there has been much progress in the past decade, observations
of the bulge and bar from the ground are inherently difficult due
to the large extinction towards the galactic center. 
Going into the infrared can remove much of the obscuring effects of the
dust, but then the limitation becomes the high stellar source density
encountered in the central regions.  WFIRST
with the reddest possible filter would offer the best means
to enable new research. Figures \ref{fig:disk_horizon} and \ref{fig:hst_VVV}
illustrate the advantages of redder filters and smaller PSFs for studies of
the inner disk and bulge.

One example of new science that would come from a wide-area $H/K$ or $J/K$ survey
of the bulge with WFIRST would be a much more detailed map of the
spatial distribution of red-clump giants (tracers of the metal rich
bulge population) than possible from the ground, definitively measuring
the orientation, length, and width of the bar and plausibly allowing one
to trace whether spiral arms originate from the ends of the bar.
A more demanding role for WFIRST, but one that could provide unique and
far-reaching new data, would be to obtain several epochs of that map
at $K$ over the 5-year WFIRST prime mission (or, better, sampling also over an
extending mission timeframe).  Such data,
when combined with the dense astrometric grid expected from Gaia, could
provide proper motions for tens to hundreds of thousands of red-clump
giants and thousands of RR Lyr stars, thereby providing kinematic
data needed to help empirically constrain the origins and evolution
of the metal poor and metal rich populations of the inner galaxy (100 km/s
at 8.5 kpc corresponds to about 2.4 mas/yr), as well as providing constraints on the
relative contributions of dark and luminous matter in the inner Galaxy.
There are plans already to use
data from the 2.8 square degrees of the microlensing campaigns 
to address these questions.   A GO program that provided $H/K$-band imaging 
of significantly more sight-lines through the bulge and inner disk,
particularly where extinction limits sensitivity at shorter wavelengths,
could not be obtained by any other facility and would greatly improve
the results from the planned microlensing survey.

Another specific project would be to obtain synoptic K band imaging of
the inner $\sim$30' of the Galaxy (four WFI fields) in order to identify all of
the RR Lyrae stars in the Nuclear Star Cluster (NSC) and Nuclear Bulge/disk.
A dozen or so RR Lyr have recently been identified in the NSC using HST
and data from the VVV survey (Dong \etal\ 2017; Minniti \etal\ 2016); however,
those surveys are very incomplete due to the extreme crowding and variable
reddening (see Figure \ref{fig:hst_VVV}).  Yet determination of the total number of RR Lyr in the NSC is vital
for constraining its formation mechanism.  Is most of the mass of the NSC from
the merger of a previous generation of globular clusters (Tremaine \etal\ 1975)
or from in-situ star-formation (Agarwal \& Milosavljevic 2011)?  A survey
with WFIRST (single epoch at J band, multi-epoch at K band) would be the
best way to identify the complete RR Lyr population of the NSC and much of
the Nuclear Bulge and thereby answer that question.   The same data could
be used to measure the proper motions of these RR Lyr and thereby determine
their kinematics.

\section{The Fragmentation Limit of Star Formation: Free-Floating
   Planets in Star-Forming Regions}

One of the goals of the WFIRST microlensing program is to determine the
number of free-floating planets in the disk of the Milky Way and to attempt
to infer their properties and origin.   WFIRST will be able to address this
same goal in a very different and complementary way by obtaining deep, multi-epoch imaging of
nearby, rich star-forming regions to identify the planetary-mass extension
of the stellar-substellar mass function and possibly to determine the
lower mass limit for objects which form as separate entities from direct
collapse of molecular-cloud cores.   Both projects together will make it possible
to determine whether the free-floating planets detected by the WFIRST microlensing survey formed
in isolation or were formed in circumstellar disks and subsequently ejected.   
The star-forming regions best surveyed for this project
should be old enough ($>$ 3-4 Myr) so that there is little on-going accretion and
few or no primordial disks remaining, both of which can alter the apparent
colors of young stars.  However, the best regions should also be young
so that the planetary mass objects are still bright and so that dynamical
evolution has not significantly affected the at-birth IMF, either through
mass segregation or preferential ejection of the lowest mass members.

Fortunately, there are a number of good, nearby target regions such as
the Scorpius-Centaurus (Sco-Cen) association or older portions of Orion.   
A $K$-band filter would
significantly improve the ability of WFIRST to address this science because
the planetary mass objects of most interest -- with masses 0.5 to 10 M$_{\rm Jup}$ --
should have temperatures in the range 1200-2000 K at ages of 5-15 Myr
(appropriate for Sco-Cen); they should thus have L dwarf spectral types
and be {\it very} red, with $V-K$ $\sim$ 10 mag and $J-K$ $\sim$ 2 mag or more.  
In fact, these planetary mass, very  young objects have been found to be
even redder than old L dwarfs of the same spectral type - possibly due
to enhanced dust formation at low surface gravities (Bowler \etal\ 2017).

\begin{figure}[ht]
\epsscale{1}
\plottwo{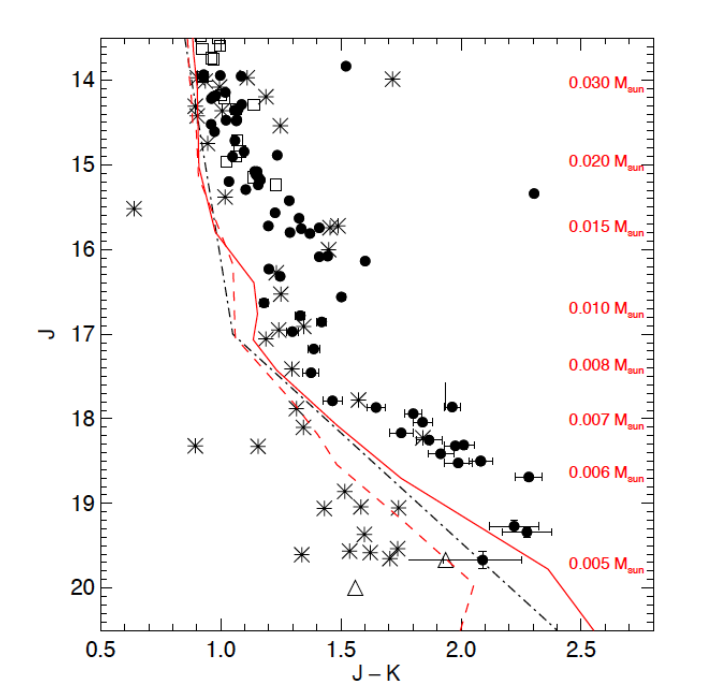}{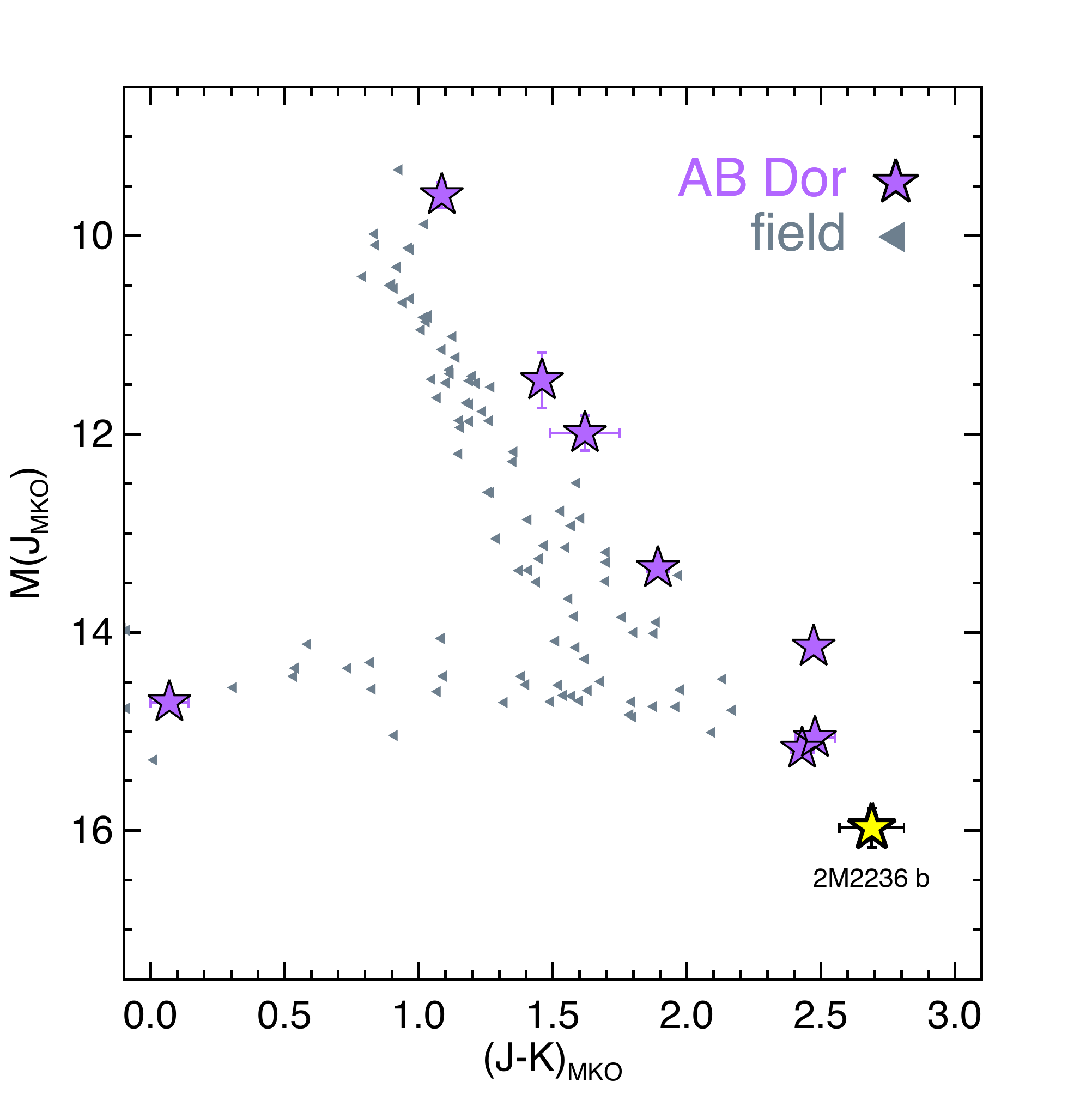}
\caption{(LHS)$J$ vs. $J-K$ color-magnitude diagram for a 13-sq-deg
region within the Upper Sco association (distance $\sim$150 pc;
age $\sim$10 Myr) obtained with the VISTA
telescope (Lodieu \etal\ 2013).   A sequence of planetary mass candidates extends to
the faint limit of the survey, with the lowest mass objects likely
being about 8 M$_{\rm Jup}$  (filled dots: likely Upper Sco members;
asterisks: photometric non-members). (RHS)  Similar color-magnitude 
diagram for low mass members of the AB Dor moving group (Bowler et al. 2017).  
The yellow star shows the location of an 11 M$_{\rm Jup}$ member of AB Dor,
which is similarly very red and faint relatively to the field star sequence.
WFIRST with a K-band filter could extend these searches to objects well
below 1 M$_{\rm Jup}$.
\label{fig:USco_planets}}
\end{figure}

Initial attempts to identify the fragmentation limit for formation of 
free-floating objects in Sco-Cen have already been made using the best available
ground-based facilities (Lodieu \etal\ 2007; 2013).  
Figure \ref{fig:USco_planets} shows the result of the deepest search
to date, which covers a 13 square degree region and appears to have identified more than a dozen planetary
mass members based on their CMD position, proper motions consistent
with Upper Sco membership, and spectroscopic confirmation for 12 of the faintest
15 objects (which shows that they are young L dwarfs) .  The faintest
objects have $K$ $\sim$ 17 mag and estimated mass of about 6 M$_{\rm Jup}$, with no
evidence that the sequence has ended.   A WFIRST survey of the entire $\sim$50-sq-deg 
Upper Sco region could go more than 5 magnitudes 
fainter at $K$, to a mass well below 1 M$_{\rm Jup}$.   With two epochs of $K$ imaging
separated by five years, and one epoch at both $J$ and $K$, such a survey
could provide much better proper motion membership data (relative to ground surveys), 
a much larger sample,
and definitively measure the fragmentation limit or show that it
is below a Saturn mass.   Because Upper Sco's proper motion is about 30 mas/yr,
the proposed WFIRST data would easily prove or refute membership. 
If these solivagant planetary mass objects have space motions of a few
km/sec relative to their stellar counterparts (because they have been
ejected from a forming disk, for example), that might also be 
within the grasp of these data (1 km/s at 150 pc corresponds to 1.3 mas/yr
motion).

\section{Detecting and Characterizing the Substellar Population of the
   Thick Disk and Halo}

\begin{figure}[ht]
\epsscale{1.15}
\plotone{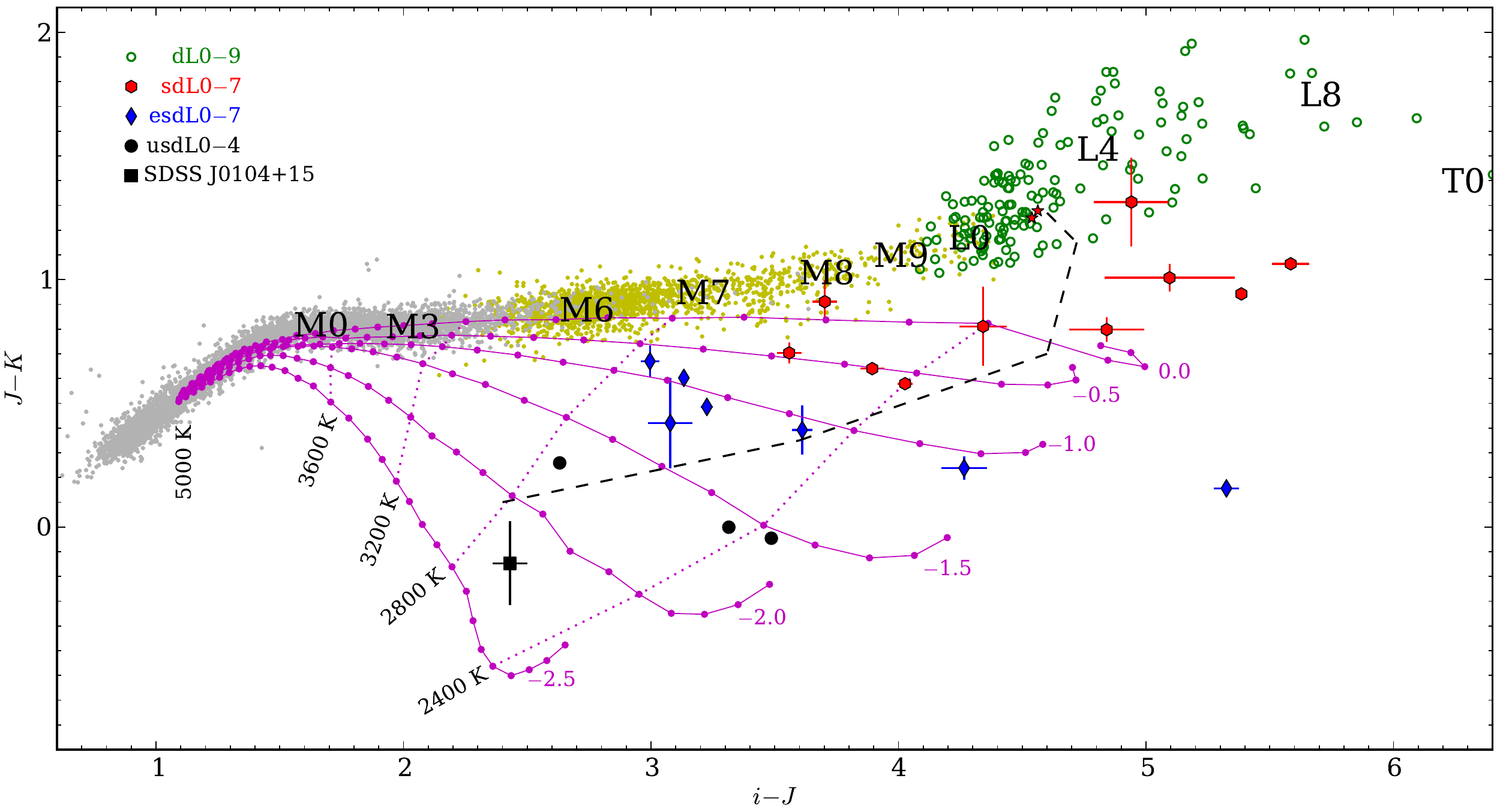}
\caption{Color-color plot for low mass stars and brown 
dwarfs, from \cite{zhang2017b}, illustrating the ability to distinguish 
metallicity classes for late-M, L, and early T dwarfs/subdwarfs using $
J-K$ color. Grey and light green dots, along with green circles 
demarcate the solar-metallicity sequence from F through L spectral types. 
Other colored points mark the locations of known late-M and L subdwarfs. 
The magenta lines represent families of curves from the BT-Settl model 
grids (\citealt{allard2014,baraffe2015}) whose metallicities (in 
magenta) and  T$_{eff}$ values (in black) are shown.   If K-band data
were added to the High-Latitude Survey, WFIRST could survey a volume
of space 5000 times larger than represented in this figure.
\label{fig:subdwarfs_wfirst}}
\end{figure}

In the past 20 years, the number of substellar objects (or, more loosely,
the number of LTY dwarfs) has gone from essentially zero to well over a 
thousand\footnote{See DwarfArchives.org.}.
As the intermediate step between stars and planets, brown dwarfs provide insights
into the fundamental physics governing the formation and evolution of both
their smaller and larger cousins.  However, because substellar objects are
both cool and small, the existing substellar census is almost entirely
representative of the Pop I disk of the Milky Way.   Only about two dozen
subdwarf L and T dwarfs (i.e. L and T dwarfs with low metallicities 
characteristic of the halo or thick disk)  have been discovered.  WFIRST
could provide the capability to not only identify a very large sample
of L and T subdwarfs, but also to sort them by temperature and metallicity
and thereby allow that census to provide fundamentally new data on 
the dependence of the star-forming process on metallicity at low mass.
Figure \ref{fig:subdwarfs_wfirst} illustrates that a filter set which includes
$J$ and $K$ band can be used very effectively to sort L dwarfs by metallicity; this
is due to collision-induced H$_2$\ absorption, which most strongly suppresses 
the $K$-band flux and thereby causes L dwarfs to have increasingly blue 
$J-K$ colors with decreasing metallicity (\citealt{borysow1997}).

The largest areal survey to date (\citealt{kirkpatrick2014,kirkpatrick2016}) used
proper motion measurements from the all-sky AllWISE processing (\citealt{cutri2013})
to identify nine L subdwarfs that, because they 
are bright and nearby, serve as the prototypes of their class.
The deepest survey for L subdwarfs to date (\citealt{zhang2017a}) used imaging
of 3000 square degrees from the UKIRT Large Area Survey and SDSS, identifying
eleven L subdwarfs down to $K$ $\sim$ 17 mag (Vega) and to distances of order 100 pc. 
If the WFIRST 2000-sq-deg HLS included a $K$-band filter -- yielding 5$\sigma$ $K$-band
detections to $K$ $\sim$ 23.5 mag (Vega) -- that survey could cover a volume five
thousand times larger than the Zhang et al. study and to distances of 2 kpc.  

The Pop I disk is rotationally supported; the halo is mostly pressure supported.
Therefore, halo stars in the solar neighborhood have streaming motions in
the direction of the disk rotation of order 200 \kms.   Thick disk stars
are intermediate in their kinematics, but still have a quite large streaming
motion relative to the Pop I disk. A single-band second epoch of the HLS taken
several years after the original HLS can essentially identify every halo and 
thick-disk star in the HLS survey area via its proper motion as long as $K$-band 
images are available for color/metallicity discrimination at one of the epochs. 
Specifically, at a distance of 2 kpc, 200 km/s corresponds to about 20 mas/yr 
if the motion is in the plane of the sky. This can be compared to the 
expectation\footnote{\url{https://conference.ipac.caltech.edu/wfirst2016/system/media_files/binaries/29/original/WFIRST_Astrometry_Spergel.pdf}}
that the HLS will be able to produce proper motions of order 0.1 mas/yr for a 
baseline of 5 years for stars brighter than magnitude 23. 

Intriguing additional science is also possible with these data. One
expects the luminosity function of old populations to show a gap between the
bottom of the stellar sequence and the most luminous substellar objects.  The 
gap should increase as the population ages.  It may be possible to measure
this gap as a function of metallicity class (surrogate for age)  in a $K$-band supplemented, dual-epoch 
HLS, providing a direct measure of the cooling timescale for high
mass brown dwarfs.

\section{Using a $K$ Filter on WFIRST To Provide A New H$_0$ Distance Ladder}

The most accurate current cosmic distance ladders to determine the
value of the Hubble constant are based, respectively, on data primarily
at $J$ and $H$ bands (Riess \etal\ 2011) and at IRAC [3.6] and [4.5]
micron bands (Freedman \etal\ 2012). The fact that the HST ladder did not use K band was not because K band is disfavored
for some astrophysical or cosmological reason, but simply because the thermal operating
temperature of HST greatly reduces its sensitivity at that wavelength.    All of the
primary initial rungs used in such distance ladders (Cepheids, RR Lyr variables, TRGB)
work better at longer wavelengths because the longer wavelengths are less affected by
reddening and metallicity variations.  If HST had the same thermal environment as now
expected for WFIRST, WFC3 would have had a K filter and the HST distance ladder would
very likely have used K band data.

In order to construct a PopII distance ladder (Beaton \etal\ 2016), accurate RR Lyrae 
distances require a well defined PL relation.  Differently than for 
Cepheids, this relation exists only at long wavelengths.
For short wavelengths the slope of the PL relation 
is just too shallow, and the intrinsic scatter too large ($>$5\%), to produce 
accurate distances. The turnaround in the slope of the PL relation happens 
in the near-IR, where the intrinsic scatter due to evolutionary effects and
temperature dependence is greatly reduced because of a narrowing in the
instability strip and because the brightness variations are mainly driven
by radius changes.  In fact, RR Lyrae distances in the K band are 
as accurate as the distances that can be derived in the L and 
M bands, with just a small penalty for the larger extinction at K compared to L or M.
The accuracy in the distances obtained in the J and H bands 
will instead be lower, not just because of the higher extinction, but also
because of the larger intrinsic scatter in the PL relation (in the V band 
there is no PL relation). The distances obtained in the K band are very 
accurate even for individual stars: this means that these variables can 
also be used to probe the geometry of the host galaxies, in addition to 
anchor the distance scale. 

Given the importance of reconciling the tension between the measured
local value of H$_0$ and the large-scale value of H$_0$ inferred from
CMB data (\citealt{bernal2016}), it seems likely that JWST will be used
to produce a new distance ladder.  
In the absence of a true K band filter, the most efficient way for JWST to 
obtain high precision distances of RR Lyrae in Local Group galaxies will 
be to use the NIRCam F200W or F277W wide filters (best for reducing 
crowding), in addition to the Spitzer/IRAC-like F356W and F444W passbands 
(where the extinction is even lower). For most of these galaxies
however, the small size of NIRCam detectors will limit these 
observations to narrow areas in each galaxy, a limitation that does not 
exist with the huge WFIRST field of view. A survey of these extended 
targets with WFIRST in the K band would allow to efficiently sample a 
much larger population of RR Lyrae than feasible with JWST, allowing to 
statistically reduce the scatter in their PL relation due to the 
metallicity dispersion within those galaxies, providing more accurate 
anchors for the JWST PopII distance scale.  By observing these fields
with both F184W and a K$_{s}$ filter, synthesized WFIRST magnitudes closely
matched to the F200W bandpass could be produced, allowing the WFIRST and
JWST data to be precisely aligned.

\section{Identifying and Characterizing Infrared Transients with WFIRST}

In addition to the survey science potential, a K-band filter on WFIRST 
would make for a powerful and unique capability for 
Time-Domain Astronomy (TDA) in the near-infrared.  It 
would go much deeper than ground-based K-band surveys, much deeper than 
LSST on a visit-by-visit basis and therefore for the same 
cadence, and much deeper (~5mag) than Spitzer. WFIRST would also offer
much better spatial resolution than LSST or Spitzer, a critical advantage 
for exploring the transient source populations in more crowded 
parts of galaxies.  Two examples of science applications of this TDA 
capability are discussed  below, but the most recent illustration comes 
from the first detection of electromagnetic counterparts to gravitational 
waves from neutron star mergers; by 10 days after its discovery, the peak 
of the SED of the NS-NS merger event GW 170817 had shifted to 2.1 $\mu$m (Pian et al. 2017).
This was predicted by Kasen et al. (2013) and Barnes \& Kasen (2013) who 
showed that the peak of the emission is beyond 1.4$\mu$m due to the large number 
of line transitions in heavy elements produced by r-process nucleosynthesis. 
More recent opacity calculations (Fryer et al. in prep.) suggest that the 
emission could be even redder and even fainter than initially predicted.

The SPIRITS survey with Spitzer has recently uncovered a population of 
transients that show predominantly (and often exclusively) infrared emission 
(see Kasliwal et al. 2017). These transients, dubbed SPRITEs, lie in the 
luminosity gap between novae and supernovae and photometrically evolve
on a wide range of timescales.    Half a dozen have been discovered in nearby galaxies
surveyed by SPIRITS out to 7 Mpc, but they would be easily detected by WFIRST
at K-band out to $\sim$100 Mpc.  The much larger sample that could be
discovered by WFIRST would provide the data necessary to firmly identify
the progenitor types and physical mechanisms which give birth
to the SPRITES.    Current possible interpretations of SPRITEs include formation 
of massive star binaries, stellar mergers, birth of stellar mass black 
holes, electron capture supernovae on extreme AGB stars etc. Since the 
emission from SPRITEs peaks in the mid-IR, with T$_{eff}$\ between 350 and 
1000 K, a redder WFIRST filter would be much more sensitive to 
discovering and characterizing these events.   Moreover, without a
K-band filter, the most interesting SPRITEs may go unidentified as such,
even if detected at optical wavelengths.

A WFIRST galactic plane survey would provide the first epoch data
needed to identify and characterize the explosive events in high-mass
star-forming regions that we have only recently begun to detect
(Bally \etal\ 2015; Caratti o Garatti \etal\ 2017; Hunter \etal\ 2017).
These events have been found in very crowded, heavily embedded regions,
requiring good spatial resolution and the longest wavelength filters possible.
Do young high mass stars grow primarily from mergers or from intermittent,
high accretion rate bursts from their circumstellar disks?  A K-band enabled 
WFIRST could best help provide the data to answer that question.  When the
next Type II supernova occurs in the Milky Way, this same WFIRST galactic
plane survey could provide the data needed to characterize the progenitor
of that event.

\begin{figure}[ht]
\epsscale{1.17}
\plottwo{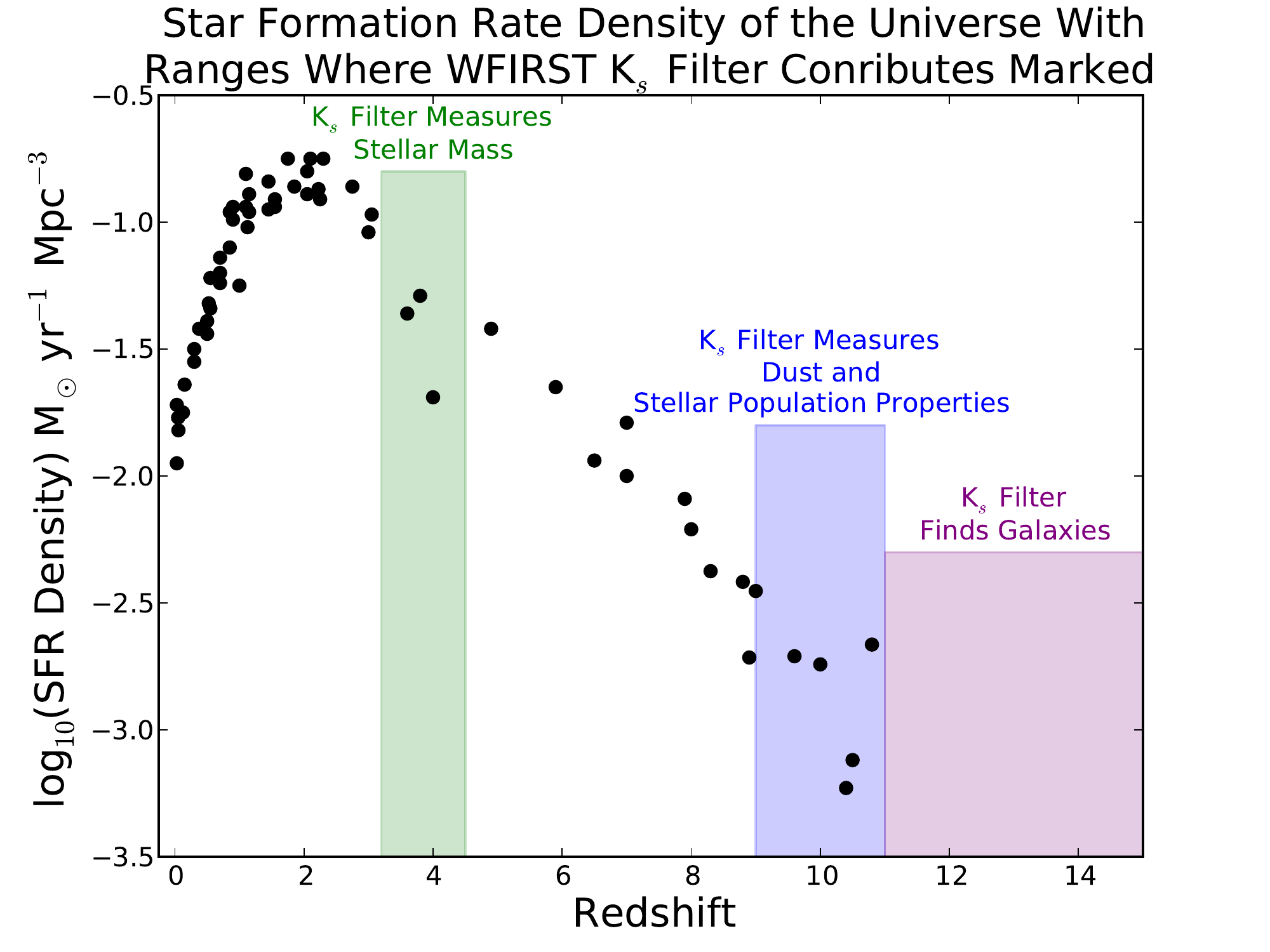}{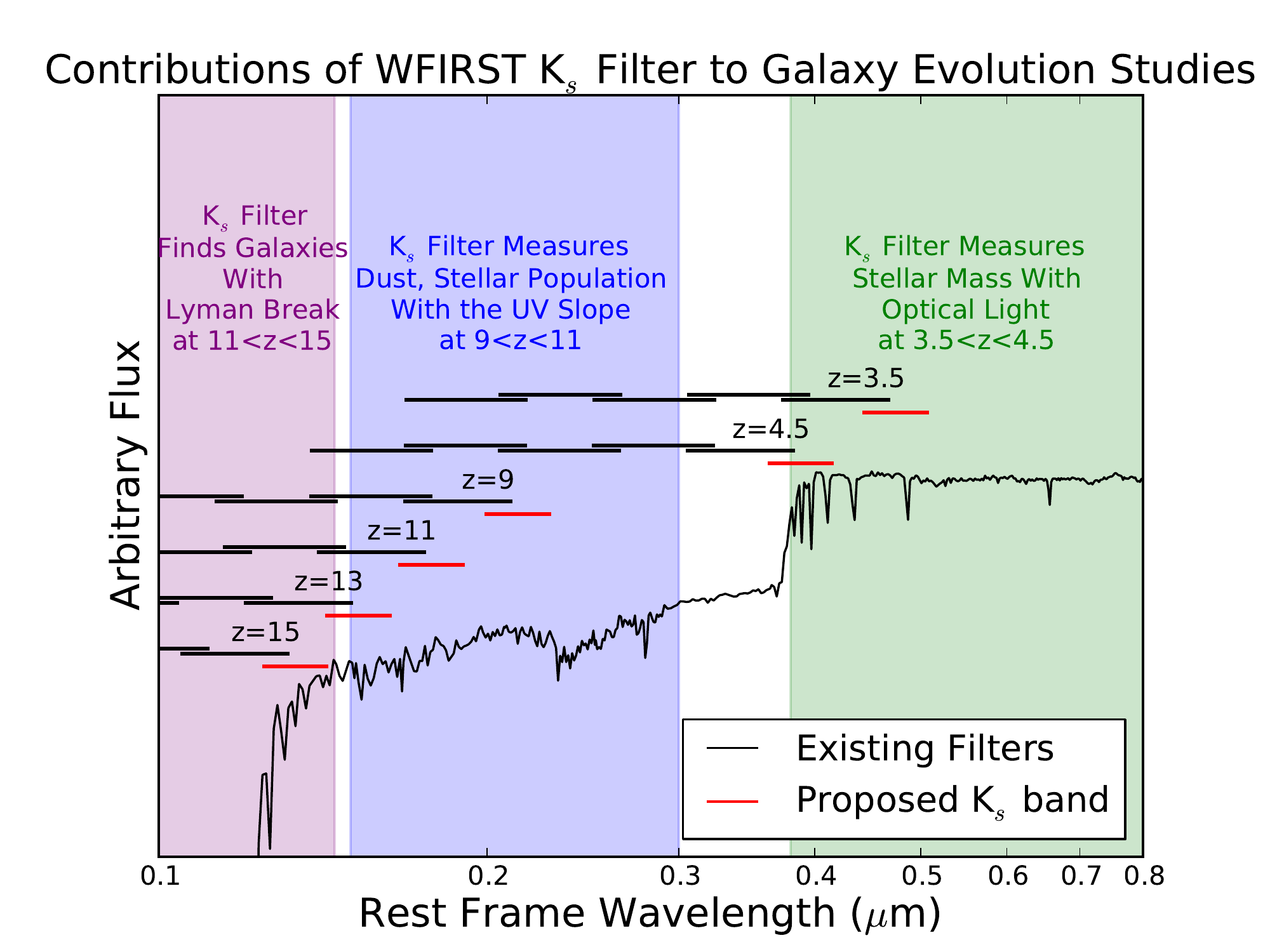}
\caption{{\bf Left:} The global history of the star formation rate density is shown using the 
compilation from Madau \& Dickinson (2014) with updated $z>8$ points from 
Ishigaki \etal\ (2018).  A K$_s$ band filter on WFIRST would make several 
key contributions to understanding the rise in the global star formation 
rate at $z>3$.  First, indicated in green, it would enable stellar mass 
estimates at $3.5<z<4.5$, which is over the peak in the global star 
formation rate density.  Second, indicated in blue, it would enable 
estimates of the ultraviolet spectral slope at $9<z<11$ which is a crucial 
period for changes in the intrinsic stellar population and dust content 
of galaxies.  Finally, indicated in purple, it would enable searches for 
$11<z<15$ galaxies which JWST does not have the survey speed to find (see Figure \ref{fig:LF}). 
{\bf Right:}The rest frame Spectral Energy Distribution (SED) of a 100Myr old 
galaxy is shown along with regions where a WFIRST K$_s$ band filter would 
contribute information.  The region of the SED probed by WFIRST filters 
at different redshifts is indicated in black for the existing Z, Y, J, H, 
F184 filters and red for the proposed K$_s$ filer.   The rest frame optical, 
indicated in green, contains information on the stellar mass and age of 
galaxies.  The 1500-3000\AA\ ultraviolet portion of the SED indicated in 
blue contains information on the intrinsic hardness of the ultraviolet 
radiation and the dust content of galaxies.  Finally, the Ly-$\alpha$ 
break at 1216\AA\ allows one to select galaxies at $11<z<15$.
\label{fig:SFRD}}
\end{figure}

\section{Improving Extragalactic Population Studies at $z$ $>$ 3}

Extragalactic surveys would benefit in several ways from extending the 
wavelength coverage to the red.  The left hand panel of Figure \ref{fig:SFRD} shows current measurements 
of the global star formation history along with regions where a WFIRST K$_s$ 
band would add critical information. First, as shown in the right hand panel of Figure \ref{fig:SFRD}, 
a K$_s$-band survey on WFIRST would extend the redshift  range for which 
stellar masses can be estimated from z $\sim$\ 3.5 
to z $\sim$\ 4.5.  This enables one to probe beyond the peak of the global star formation at 
z $\sim$\ 2-3 (Madau \& Dickinson 2014).  The combination of sensitivity 
and area that WFIRST could provide compared to ground based and existing Spitzer 
surveys would enable studies of early dwarf galaxy formation linked to 
dark matter formation via galaxy clustering measurements 
(Lee et al. 2012, 2009, Finkelstein et al. 2015).  This co-measurement of 
clustering and galaxy mass provides a strong constraint on the duty cycle 
and possible star formation histories in the early universe 
(Lee et al. 2009, Finkelstein et al. 2015). 

Second, the addition of a longer-wavelength filter would significantly improve the selection 
efficiency of 9 $<$ z $<$ 11 galaxies and enable measurements of their rest-frame UV spectral 
slopes, which are very sensitive to their evolving stellar populations and dust content
(Finkelstein et al. 2012, Madau and Dickinson 2014, Figure \ref{fig:SFRD}).  

\begin{figure}[ht]
\epsscale{1.15}
\plotone{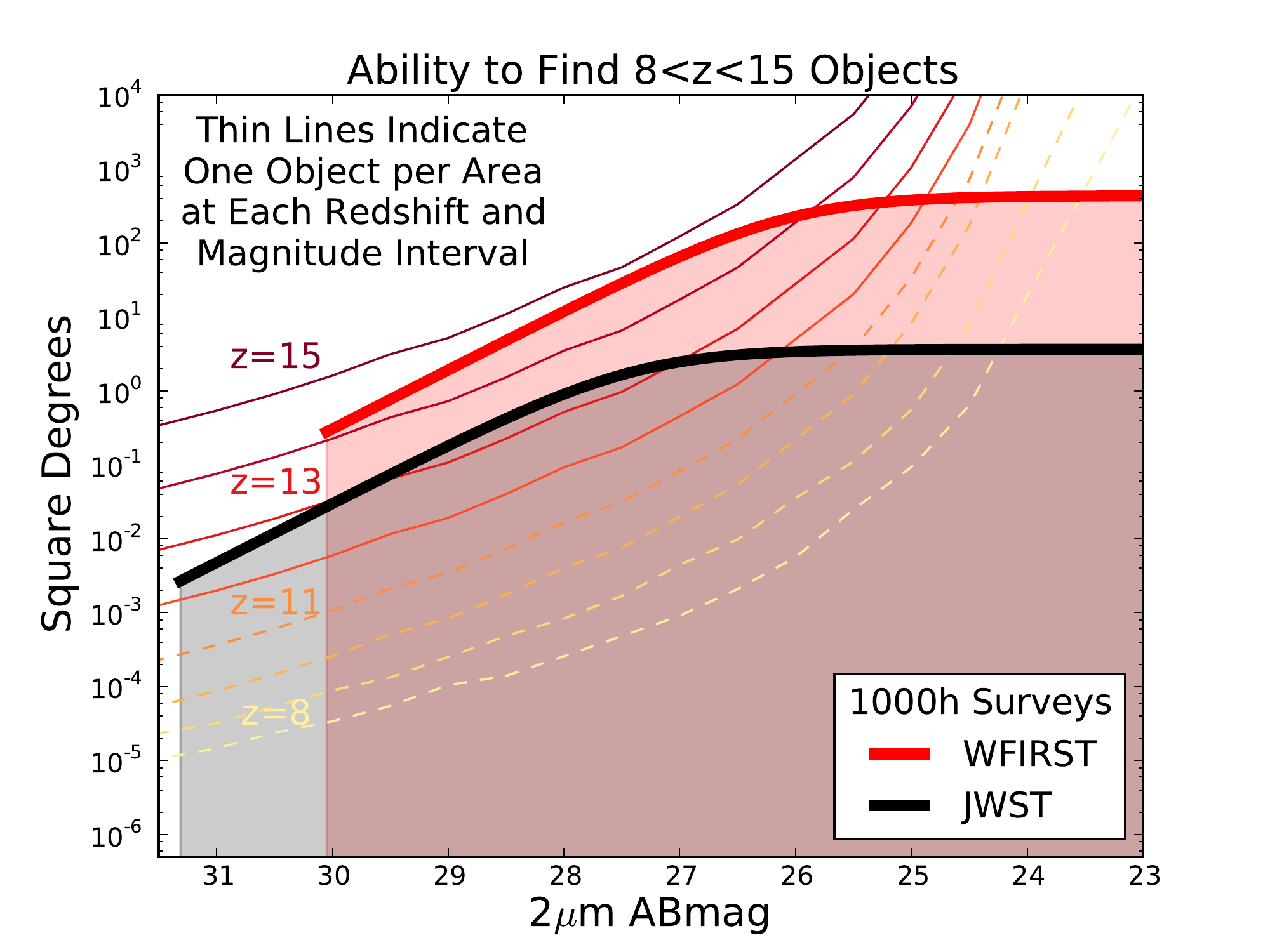}
\caption{The thick red and black lines show the depth and area reachable 
by WFIRST and JWST in 1000h respectively.  A given 1000h survey would be 
one point on the line and probe areas and depths smaller than the point.  
For WFIRST we assumed a survey in 5 filters (Y,J,H,F184W, K$_s$) and used 
the current overhead estimates provided by C. Hirata (private communication).  
For JWST we assumed a 3 filter (F115W, F150W, F200W) NIRCAM survey with 
the actual overheads given by the Astronomers Proposal Tool (APT) excluding the 0.5h slew to field overhead.  Galaxies were assumed to be $0.1^{\prime\prime}$ FWHM gaussians when estimating sensitivity. The 
area where one object is expected in a given survey per magnitude and 
redshift interval is indicated with thin dashed lines for redshift ranges accessible with the existing WFIRST filter complement and solid lines for redshift ranges only reachable
if WFIRST had a K$_s$ band filter.  With these filters JWST would not be able to select $z<9$ galaxies due to the lack of Y band data. The lines are based on models of measured data at $8<z<10$ from 
Yung \etal\ (2018).  At $11<z<15$ the $z=10$ Yung \etal\ (2018) estimate is scaled by the density evolution of dark matter given in Mashian \etal\ (2016). The error on the density in these models is $\pm0.2-0.5$ dex due to a combination of current measurement error, cosmic variance, and the assumptions in the extrapolations.  
\label{fig:LF}}
\end{figure}

Finally, as shown in Figures \ref{fig:SFRD} and \ref{fig:LF} the longer  
wavelength lever arm could also be used to find 11 $<$ z $<$ 15 galaxies 
if they exist (Madau and Dickinson 2014), a task that only a many square 
degree WFIRST 
deep field could address due to the limited survey speed of JWST 
(Yung \etal\ 2018, Mashian et al. 2016).  Such a deep survey 
would also find heavily obscured super-starbursts at z $>$ 4 
(Caputi et al. 2012, Wang et al. 2012), which are also very faint 
and rare on the sky.

\begin{figure}[ht]
\epsscale{1.15}
\plotone{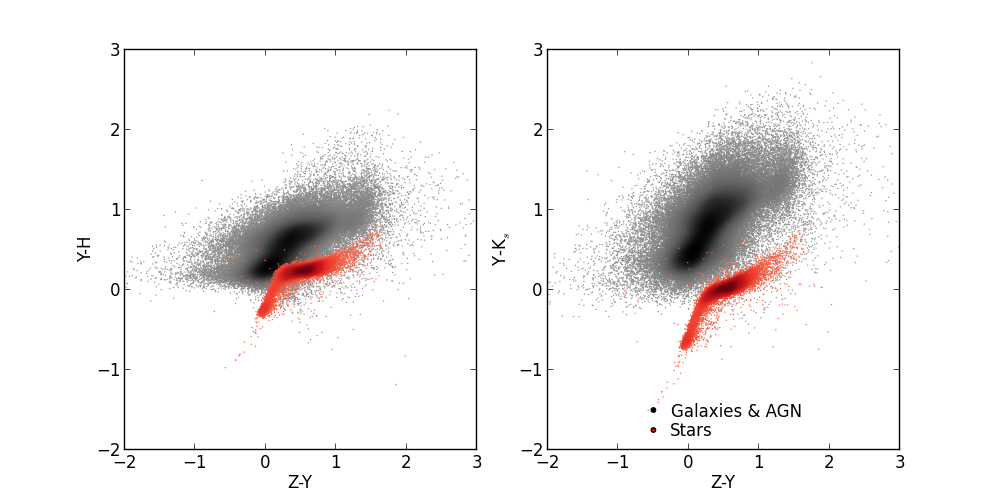}
\caption{Adding a K$_s$ band to WFIRST (right), increases the separation
between stars (red) and other objects (black) compared with the H band
(left) allowing for clear differentiation between stars and other
objects.  Actual data from the COSMOS (Laigle et al. 2016) catalog are
shown using HST F814W as a stand in for WFIRST Z band and Ultra-Vista
Y, H, and K$_s$ bands.  The star/galaxy separation in this plot is based on HST
F814W morphology along with the Laigle et al. (2016) catalog.
\label{fig:star_galaxy}}
\end{figure}

Another key advantage of a red filter for extra-galactic science is the 
ability to find and differentiate quasars from stars and galaxies to low 
luminosities, which is difficult without longer-wavelength IR data 
(Masters et al. 2012, Figure \ref{fig:star_galaxy}).  It appears that the population of very massive 
black holes that form quasars is dropping rapidly at z $>$ 6 
(e.g. Banados et al. 2016; Mazzucchelli \etal\ 2017).  The ability to measure the upper envelope 
of the quasar luminosity function to z $\sim$\ 10 combined with fainter quasars 
at lower redshifts will place strong constraints on early black hole 
formation models.

\section{Characterizing Water Ice on Distant Solar System Objects with WFIRST}

\begin{figure}[ht]
\epsscale{0.75}
\plotone{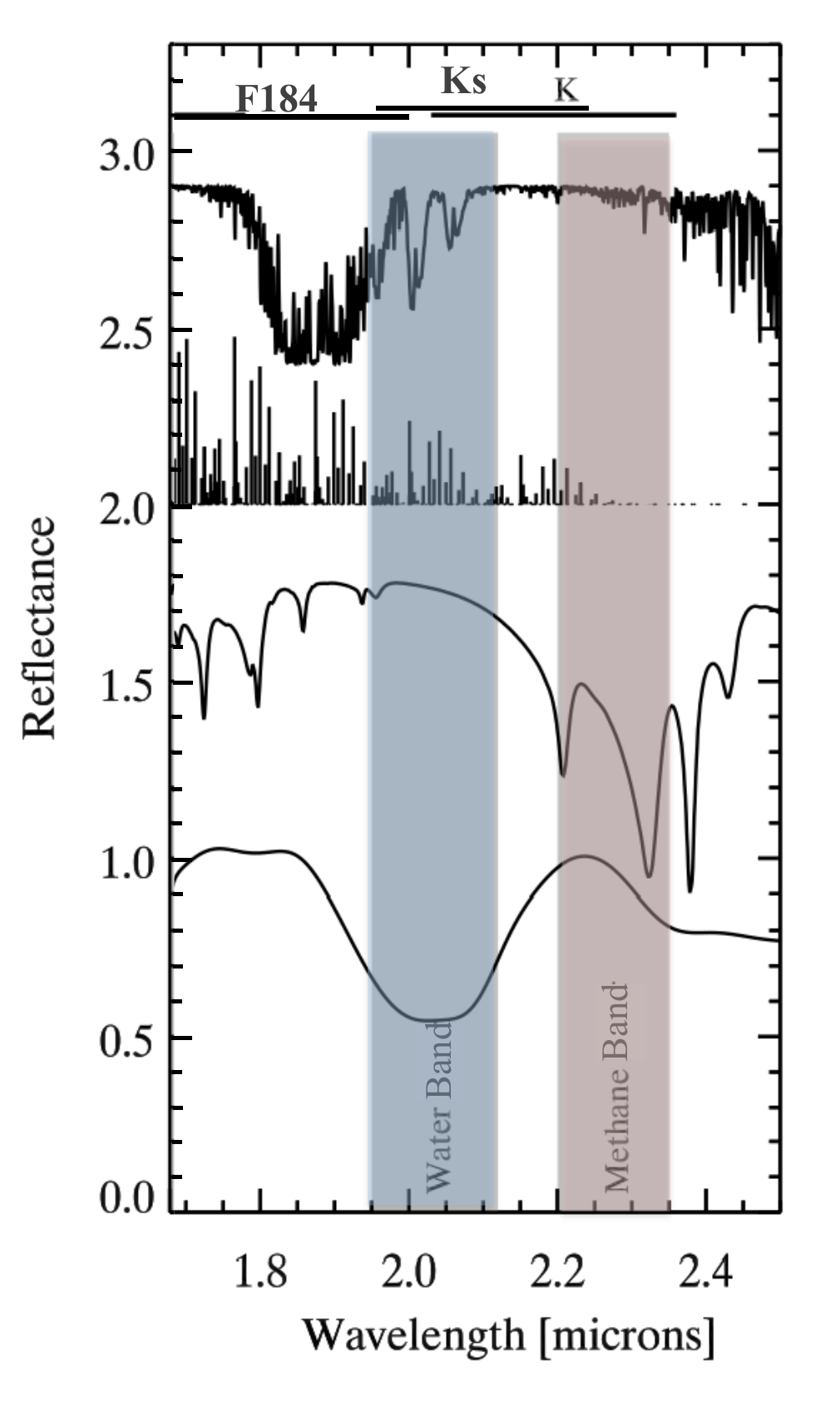}
\caption{As adopted from Trujillo et al. (2011), a Ks-band filter is dominated by strong water-absorption, though even a K-band filter would be diagnostically useful. 
\label{fig:IceTNOwfirst}}
\end{figure}

A K$-$band filter, particularly a K$_s$$-$band filter, can facilitate searches for 
volatiles on the surfaces of the outer solar system small bodies (Figure \ref{fig:IceTNOwfirst}; \citealt{truj2011}). 
Both planned and possible GO large-area surveys (e.g. HLS) will yield imaging data that will 
contain many small bodies, and those with apparent slower non-sidereal motions, i.e. the 
outer solar system small bodies, Centaurs and Trans-Neptunian Objects (TNOs), will have
only slightly elongated images in $\sim$ 3 minute medium-deep exposures, thus 
minimizing confusion and read-noise effects due to trailing. These TNOs and Centaurs
are also the most likely bodies to have detectable surface water ice,  since most or
all of them will have formed 
beyond the ``snow line" where the water-ice sublimation rate rapidly drops off
(c.f.  \citealt{dones2015}). Furthermore, because within that population the larger objects are more likely 
to have retained their surface volatiles since their formation (\citealt{schall2008}), it is 
likely that it is the brighter objects which will 
manifest water ice features.

While not exclusively within the water-ice band at 2-$\mu$m, a strong absorption feature 
would dominate the Ks band, relative to the Y, J and F184 bands  which should be
unaffected by strong water-ice absorption.  Adding Ks-band imaging to such wide-area surveys
would allow identification of potentially hundreds of icy outer-solar system bodies, 
with the brightest TNOs or Centaurs being promising targets for follow-up spectral 
observations with JWST or other future facilities. The surveys themselves would place 
meaningful constraint on the ubiquity of water ice among the TNO and Centaur populations. 

\newpage
\section{Summary and Conclusions}

The discovery space available for a redder WFIRST filter 
is huge, and there is no forseeable alternative source for
a comparable $K$-band deep/wide survey.   We are confident the 
community would find a myriad of ways to exploit that discovery space.

\clearpage

\appendix

Table~\ref{tab:Ks_data}  provides
estimated WFIRST sensitivities for a filter similar to the 2MASS K$_s$
filter, for three possible operating temperatures (J. Kruk, private communication).
\vskip0.3truein

\begin{deluxetable*}{lcccc}
\tabletypesize{\scriptsize}
\tablecolumns{5}
\tablewidth{0pt}
\tablecaption{S/N Achieved as a Function of $K_s$-band mag in an HLS Survey\label{tab:Ks_data}}
\tablehead{
\colhead{$K_s$ mag} & 
\colhead{$K_s$ mag} & 
\colhead{T=260K} &
\colhead{T=270K} &
\colhead{T=284K} \\
\colhead{(AB)} & 
\colhead{(Vega)} & 
\colhead{} &
\colhead{} &
\colhead{} 
}
\startdata
\hline \\
18.00 & 16.15  & 946.1 &  930.9  &  879.5 \\
18.50 & 16.65 & 754.4 & 727.0  & 668.0 \\
19.00 & 17.15 & 584.7 & 562.7 & 497.9 \\
19.50 & 17.65 & 455.5 &  430.2 & 362.6 \\
20.00 & 18.15 & 351.3 & 323.3 & 257.3 \\
20.50 & 18.65 & 267.3 & 237.8 & 178.0 \\
21.00 & 19.15 & 199.6 & 170.6 & 120.2 \\
21.50 & 19.65 & 145.7 & 119.2 & 79.6 \\
22.00 & 20.15 & 103.7 & 81.3 & 51.9 \\
22.50 & 20.65 & 71.9 & 54.2 & 33.5 \\
23.00 & 21.15 & 48.6 & 35.6 & 21.4 \\
23.50 & 21.65 & 32.2 & 23.0 & 13.6 \\
24.00 & 22.15 & 21.1 & 14.8 & 8.7 \\
24.50 & 22.65 & 13.6 & 9.4 & 5.5 \\
25.00 & 23.15 & 8.7 & 6.0 & 3.5 \\
25.5 & 23.65 & 5.5 & 3.8 & 2.2 \\
\enddata
\tablenotetext{a}{Assumes five exposures with a total observing
time of 868 s.  For $K_s$ band, m(AB) $-$ m(Vega) = 1.85.  Zodiacal level
set at 1.44 times the minimum.} 
\end{deluxetable*}

\end{document}